\documentclass[a4paper]{JHEP3}
\usepackage{amsmath,amssymb,calc,latexsym,mcite,bbm}
\usepackage{graphicx,epsfig}

\title{Deconstructing Non-Abelian Gauge Theories\\at One Loop}

\author{Zoltan Kunszt$^{\dagger,*}$, Andreas Nyffeler$^{\dagger}$ and Martin
Puchwein$^{\dagger,*}$\\

$^\dagger$Institute for Theoretical Physics, ETH, CH-8093
Z\"urich, Switzerland\\
$^*$Kavli Institute for Theoretical Physics, University of
California, \\
\phantom{$^*$}Santa Barbara, CA 93106, USA\\
\vspace{-1mm} E-mail: \email{kunszt@phys.ethz.ch},
\email{nyffeler@phys.ethz.ch}, \email{puchwein@phys.ethz.ch} }

\abstract{Deconstruction of 5D Yang-Mills gauge theories is
studied in next-to-leading order accuracy. We calculate one-loop
corrections to the mass spectrum of the non-linear gauged
$\sigma$-model, which is the low energy effective theory of the
deconstructed theory. Renormalization is carried out following the
standard procedure of effective field theories. The relation
between the radius of the compactified fifth dimension and the
symmetry breaking scale of the non-linear $\sigma$-model is
modified by radiative corrections. We demonstrate that  one can
match
 the low lying spectrum of the gauge boson masses of the effective 4D
gauged non-linear $\sigma$-model to the Kaluza-Klein modes of the
5D theory at one-loop accuracy.

}

\preprint{NSF-KITP-04-26\\
hep-ph/0402269 \\
February 26, 2004}

\begin{document}
\section{Introduction}
\label{sec:intro} In the last few years the idea that at short
distances the fundamental interactions may be best described by
field theories with one time and more than three space dimensions
had an enormous impact on particle physics model building
\cite{Arkani-Hamed:1998nn, Antoniadis:1990ew}. It has been used
in attempts
 to solve the gauge hierarchy problem, to
cure shortcomings  of GUT models, to build models for the flavor
structure and to construct new types of gauge and supersymmetry
breaking mechanisms just to mention a few of the numerous
applications (see \mcite{Giudice:2003pj,*Barbieri:2003dd} and
references therein). In the simplest physical picture the
additional space dimensions are compact\footnote{Infinitely large
extra dimensions with warped geometries have also been suggested
\cite{Randall:1999vf}.}
 and at distances large compared to the size of these dimensions
the physics appears to be four dimensional. At  energies higher
than the inverse radius of the compact dimensions the low lying
Kaluza-Klein states can be excited. At even higher energies the
full nature  of higher dimensional physics will  be revealed. The
difficulty is, however, that higher dimensional theories are not
renormalizable and therefore ill defined as quantum field
theories. They can only be used  as effective field theories
(EFTs) defined with a cutoff and limited range of
validity.\footnote{Field theories in five
 or six dimensions may exist around a non-gaussian ultra-violet
 fixed point\cite{Seiberg:1996bd,Seiberg:1997qx,Gies:2003ic}.}
 At energies close to the cutoff they
become typically strongly coupled and
 their  predictions will depend strongly on the cutoff scheme.
These difficulties may be resolved by an ultra violet (UV)
completion which respects all the important symmetries of the
quantum field theory (Lorentz invariance, gauge invariance, SUSY,
etc). For example, 5D supersymmetric Yang-Mills theories could be
embedded in various ways into string theory or M-theory
\cite{Iqbal:2002ep}. However, a pure QFT embedding is also
possible \cite{Arkani-Hamed:2001ca,Hill:2000mu}, in particular
 extra dimensions can be
generated dynamically from anomaly free asymptotically free gauge
theories. The symmetries and particle content of the theory are
represented by ``moose'' diagrams in ``theory space'' with sites
and links. Locality and particle propagation in the extra
dimensions are maintained due to the nearest neighbor
interactions.
 In simple models at low energies
the link fields are non-linear $\sigma$-model Nambu-Goldstone
boson fields, they are mostly higgsed to give  massive gauge
bosons with a spectrum which may match the Kaluza-Klein spectrum
of compactified extra dimensions. Therefore within 4D models one
can reproduce an apparently higher dimensional mechanism. In the
simplest cases the ``moose'' diagrams lead to discretized actions
of extra dimensional gauge theories where only the extra
dimensions are latticized. Deconstruction by means of an EFT can
be viewed as a gauge-invariant regulator of the 5D
theory\footnote{An overview of other possible regularizations of
higher dimensional field theories can be found in
\cite{Ghilencea:2003xy}.}, where the lattice spacing corresponds
to the cutoff of the higher dimensional theory. These
constructions have been proven to be powerful also for physics
beyond the standard model with a few sites and without any higher
dimensional interpretation (little Higgs models)
\cite{Arkani-Hamed:2001nc}.

The mechanism of dimensional deconstruction has been demonstrated
at tree-level by matching the low lying spectra and parameters of
the 4D and higher dimensional theory. The correspondence has been
established precisely and it has been shown that this also holds
for supersymmetric theories \cite{Csaki:2001em,Hebecker:2002vm},
even in the non-perturbative regime \cite{Csaki:2001zx}. It is of
interest to investigate this correspondence also beyond tree
level. Corrections in the mass spectrum may have important
phenomenological consequences for the higher dimensional gauge
models. In some of these models, one-loop corrections may control
the electroweak symmetry breaking. In recent calculations of the
one-loop corrections of the gauge boson masses it was found that
the corrections have a well defined finite part which vanishes in
the limit of a large compactification radius
\cite{vonGersdorff:2002as,Cheng:2002iz,Puchwein:2003jq,DaRold:2003yi}.
The divergent contributions, however, are the same as in the
uncompactified theory. In the case of $U(1)$ gauge theory using
the linearized $\sigma$-model \cite{Falkowski:2003iy} for
deconstruction, these results have recently been reproduced.

In this paper we calculate one-loop corrections in the low energy
EFT, which is described by a 4D non-linear $\sigma$-model with
non-abelian gauge symmetry, and perform renormalization by adding
counter terms in chiral perturbation theory. To recover the result
of the 5D compactified gauge theory one should define the
compactification radius. We show  how the formula for the lattice
spacing obtained by the tree level matching is modified by
radiative corrections. The one-loop corrections obtained in terms
of renormalized quantities of the 4D theory can then be
reinterpreted in terms of 5D gauge coupling and compactification
radius.

In Section~2 we review the 4D $SU(k)$ gauged non-linear $
\sigma$-model together with some next-to-leading order terms in
the low energy expansion of Goldstone boson interactions. In
Section~3 we present our results for the gauge and the pseudo
Goldstone boson mass and discuss the next-to-leading order
matching to the 5D gauge theory results. Finally
 the appendices are devoted to some technical details on
propagators, the construction of the $\mathcal{O}(p^4)$ Lagrangian
and on the individual
 diagrammatic contributions.

\section{Effective field theory}
\label{sec:EFT}

A 5D $SU(k)$  gauge theory with one dimension compactified on
$S_1$ can be deconstructed using a 4D EFT that approximates the
spectrum of the first $N$ Kaluza-Klein modes
\cite{Arkani-Hamed:2001ca,Hill:2000mu,Arkani-Hamed:2001nc}. The
effective Lagrangian is obtained by using a non-linear realization
of the gauge symmetry, employing well-known techniques from chiral
perturbation theory for low-energy
QCD~\cite{Weinberg:1979kz,Gasser:1984yg} and for strongly
interacting electroweak symmetry breaking
sectors~\mcite{Appelquist:1980vg,*Longhitano:1980iz,*Longhitano:1981tm,Appelquist:1993ka,Nyffeler:1999ap}.
The effective Lagrangian is organized in a power series in small
momentum $p\!\ll\!\Lambda\!\sim\!4 \pi v$
\begin{equation}
\mathcal{L}_\mathrm{eff} = \mathcal{L}_2 + \mathcal{L}_4 + \ldots\
,\mathrm{where}\quad
 \mathcal{L}_2 = \mathcal{L}_{\mathrm{kin}} +
\mathcal{L}_{\mathrm{g.f.}} + \mathcal{L}_{\mathrm{gh}}\
.\label{lag2}
\end{equation}
The kinetic terms are
\begin{equation} \label{lag2_kin} \mathcal{L}_{\mathrm{kin}} = \sum_{p=1}^N\Big\{ v^2 \mathrm{Tr}\big[\big(D_\mu U_p\big)^\dagger
D^\mu U_p\big] -\frac{1}{2} \mathrm{Tr}\big[ F_{\mu\nu}^p
F_p^{\mu\nu}\big]\Big\}\ , \end{equation}
where the covariant derivative is defined by\footnote{We use the following
  conventions for the generators $T^a (a = 1, \ldots, k^2 -1)$ of $SU(k)$:
  $\left[ T^a , T^b \right] = i f^{abc} T^c$, $\mathrm{Tr} \left[ T^a T^b
  \right] = \frac{1}{2} \delta^{ab}$, as well as $f^{acd} f^{bcd} = C_2(G)
  \delta^{ab}.$}
\begin{equation} \label{cov_deriv} D_\mu U_p = \partial_\mu U_p- i g A^p_\mu U_p + i g U_p A^{p+1}_\mu , \qquad A^p_\mu =
A^{p,a}_\mu T^a. \end{equation}
The unitary matrices $U_p$ are given in terms of the pseudo
Goldstone bosons
\begin{equation} U_p = \mathrm{e}^{i G_p/v}, \qquad G_p = G_p^a T^a, \end{equation}
and transform linearly under gauge transformations
\begin{eqnarray} U_p \to V_p U_p V_{p+1}^\dagger, \qquad A_\mu^p \to V_p A_\mu^p V_p^\dagger + \frac{i}{g}  V_p
\partial_\mu V_p^\dagger . \label{eq:gaugetrans}\end{eqnarray}
\FIGURE[ht]{
\includegraphics[scale=1]{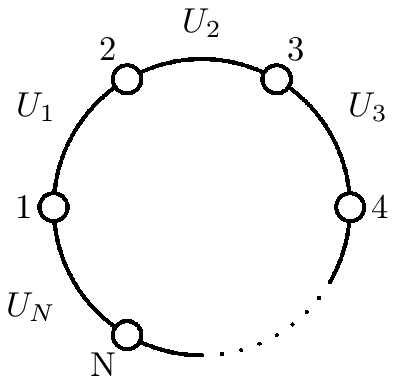}
\caption{Moose diagram} \label{fig:moose}} The model can
conveniently be described by the moose diagram in
Figure~\ref{fig:moose}. We will assume throughout that $N$ is
even.
 The Lagrangian
$\mathcal{L}_{\mathrm{kin}}$ is a gauged non-linear $\sigma$-model
with $N$ gauge groups, with equal gauge coupling $g$, provided we
impose discrete translational invariance around the circle,
$p\!\to\!p + 1$. This model can also be viewed as a 5D pure-gauge
theory formulated on $\mathbb{R}_4\times S_1$, where the
compactified dimension is latticized~\cite{Bardeen:1976tm}. Note
that the gauge fields $A^\mu_p$ live on the sites, whereas the
$\sigma$-model fields $U_p$ are associated with the link between
two adjacent lattice sites. The correspondence is established by
identifying the Goldstone boson field with the fifth component of
the higher dimensional gauge field. At tree-level the
identification is as follows (note that this relation receives
corrections at the loop-level)
\begin{equation}
\label{eq:treeid} G_p/v=-a g A_5^p\ ,\quad a^2=\frac{1}{g^2 v^2}\
,\quad a=\frac{2\pi R}{N}\ .
\end{equation}
Here $a$ denotes the lattice spacing. The dimensionful gauge
coupling of the 5D theory is given by $g_5^2=a g^2$. From a
physical point of view $a$ is related to the cutoff of the 5D EFT
($\Lambda_{\mathrm{5D}}\sim a^{-1}$). Since for this theory to be
valid it is necessary that $g_5^2 \Lambda_{\mathrm{5D}}<1$, it is
automatically ensured that in the deconstructed theory one is in
the perturbative regime.

By choosing an $R_\xi$-type gauge-fixing term
\begin{equation}
\mathcal{L}_{\mathrm{g.f.}}=-\frac{1}{\xi}\sum_p\mathrm{Tr}\big[
\partial_\mu A^\mu_p+\xi g v(G_p-G_{p-1})\big]^2
\end{equation}
we can remove the kinetic mixing term between the gauge and the
Goldstone boson fields. The corresponding Faddeev-Popov ghost
Lagrangian can then be constructed in the usual way, although some
care has to be taken due to the non-linear realization of the
gauge symmetry. Following \cite{Montvay:1994cy}, it can be written
in closed form as follows (see also
\mcite{Herrero:1994nc,*Espriu:1995ep})
\begin{eqnarray}
\mathcal{L}_{\rm gh} & = &  \sum_p\Big[ \partial_\mu\, \bar{c}_p^{\,a}\,
\partial^\mu\, c_p^a+ g f^{abc}\partial_\mu
\bar{c}_p^{\,a} A_p^{\mu,b} c_p^c \nonumber \\
& & \phantom{\sum_p} -\xi g^2 v^2
(\bar{c}_{p+1}^{\,a}-\bar{c}_p^{\,a})\big(
E^{-1}(\overline{G}_p/v)^{ab} c_{p+1}^b -
E^{-1}(-\overline{G}_p/v)^{ab} c_p^b\big)\Big]
\end{eqnarray}
where the matrix valued function $E(y)$ has the form
\begin{equation} E^{-1}(y) = \frac{y}{1 - e^{-y}} = 1 + \frac{y}{2} + \frac{y^2}{12} + \mathcal{O}\left(y^4\right)\quad
\mathrm{and}\quad \overline{G}_p^{\,ab}=f^{abc} G_p^c \ .
\end{equation}
Expanding up to fourth order in the fields, one obtains explicitly
\begin{eqnarray}
\mathcal{L}_{\rm gh}  &=&  \sum_p \Big[ \partial_\mu \bar{c}_p^a
\partial^\mu c_p^a+ g f^{abc}
\partial_\mu \bar{c}_p^a A_p^{\mu,b} c_p^c -\xi g^2 v^2 \big(\bar{c}_{p+1}^a-\bar{c}_p^a\big)\big(c_{p+1}^a
-c_p^a\big)\nonumber\\&&\phantom{\sum_p}+\frac{\xi g^2 v}{2}
f^{abc} \big(\bar{c}_{p+1}^a-\bar{c}_p^a\big) G_p^b
\big(c_{p+1}^c+c_p^c\big)\nonumber\\&&\phantom{\sum_p}- \frac{\xi
g^2}{12} f^{abc}f^{cde}\big(\bar{c}_{p+1}^a-\bar{c}_p^a\big)G_p^b
G_p^d\big(c_{p+1}^e -c_p^e\big)+(\mathrm{higher\ terms})\Big]\ .
\end{eqnarray}
Note that there appear new, quartic interaction vertices between Goldstone bosons and ghosts, compared to the case
where the symmetry is linearly realized.

To diagonalize the kinetic Lagrangian one expands the field in
terms of Fourier-modes (see Appendix~\ref{sec:fourier}). The mass
spectrum is given by
\begin{equation}
m_n^2=4 g^2 v^2\sin^2 \frac{n \pi}{N}
\end{equation}
for $A_\mu$ and by $\xi m_n^2$ for the Goldstone boson. For large
$N$, which means $R^{-1}\!\ll\!\Lambda_{\mathrm{5D}}$, the masses
of the lowest lying modes become approximately independent of $N$
and one gets an identical spectrum as in 5D Kaluza-Klein theories.
\begin{equation}
 m_n^2 \simeq \frac{n^2}{R^2}\ ,\qquad
\mathrm{for}\ n\ll N .
\end{equation}

Although the non-linear $\sigma$-model Lagrangian $\mathcal{L}_2$
in Eq.~(\ref{lag2_kin}) by itself is non-renormaliz\-able,  it is
possible to remove all divergences in the EFT order by order in
the low-energy expansion, since only a finite number of terms
appear up to a given order in $p$. For instance, at
$\mathcal{O}(p^2)$ there are only tree-level contributions from
$\mathcal{L}_2$, whereas at $\mathcal{O}(p^4)$ one has to consider
tree-graphs from $\mathcal{L}_4$ and one-loop graphs with vertices
from $\mathcal{L}_2$. The Lagrangian $\mathcal{L}_4 = \sum_i c_i
{\cal O}_i$ then contains higher order operators ${\cal O}_i$
which are multiplied by the bare low-energy constants\footnote{We
will use dimensional regularization with $d =
  4 - \epsilon$ which preserves chiral and gauge invariance. Furthermore
  contributions to the action from the integration measure $[\mathrm{d} U]$ in
  the path integral  vanish in dimensional regularization. }
\begin{equation}\label{eq:loenco}
c_i = \delta_i \mu^{-\epsilon}
\left(\frac{2}{\epsilon}-\gamma+\log 4\pi\right) +
c^\mathrm{r}_i(\mu)\end{equation} Each constant contains a
divergent part which absorbs the divergence that arises from loops
with vertices from $\mathcal{L}_2$. The coefficient $\delta_i$ of
the divergent term is unambiguously given within the EFT
framework. The finite, scheme- and scale-dependent renormalized
low-energy constants $c^\mathrm{r}_i(\mu)$ encode the different
underlying theories, e.g.\ a strongly interacting gauge theory,
like QCD, or some linear $\sigma$-model with a heavy Higgs boson.
The renormalization group running of $c^\mathrm{r}_i(\mu)$ is
again completely fixed within the EFT. In principle, these
low-energy constants can be derived by performing a matching
between effective and full theory at low energies. In practice,
e.g.\ for a strongly interacting case, one has to determine them
from experiment.

There is a well known procedure to construct all possible terms in
the effective Lagrangian at a given order in the  low-energy
expansion~\mcite{Arkani-Hamed:2001nc,Gasser:1984yg,Appelquist:1980vg,*Longhitano:1980iz,*Longhitano:1981tm,Appelquist:1993ka,Nyffeler:1999ap,Urech:1995hd}.
First of all, it is useful to assign the following chiral
dimensions to the different fields
\begin{equation} U_p=\mathcal{O}(1)\ ,\qquad G_p=\mathcal{O}(1)\ ,\qquad D_\mu
  U_p=\mathcal{O}(p)\ ,\qquad
A_\mu^p=\mathcal{O}(1)\ . \end{equation}
These assignments determine the order of individual terms in the
chiral Lagrangian by power counting. In the present case, one has
to take into account that there is a larger global chiral symmetry
if we switch off all the gauge couplings in Eq.(\ref{cov_deriv}).
The Lagrangian in Eq.(\ref{lag2_kin}) is then invariant under the
following $2N$ global chiral transformations of the $N$ non-linear
$\sigma$-fields $U_p$
\begin{equation} U_p\rightarrow e^{i\xi^p_L} U_p e^{-i\xi^{p+1}_R}, \quad e^{i\xi^p_{L,R}} \in SU(k), \quad \xi^p_{L,R} =
\xi^{p,a}_{L,R} T^a . \end{equation}
From comparison with Eq.(\ref{eq:gaugetrans}) one sees that at
each point $p$ the gauge group is just the diagonal of two
adjacent chiral symmetry groups generated by $\xi_L^p$ and
$\xi_R^p$. We therefore require that under global chiral
transformations the gauge fields transform as follows
\begin{equation} A_\mu^p\rightarrow e^{i(\xi_L^p+\xi_R^p)/2} A_\mu^p
e^{-i(\xi_L^p+\xi_R^p)/2}\ .\end{equation} Half of the chiral
symmetries are broken by the gauge interaction and every breaking
is accompanied by a factor of the gauge coupling constant $g$. To
keep track of the chiral symmetry breaking we introduce
matrix-valued spurion fields\footnote{During the preparation of
this manuscript \cite{Hirn:2004ze} appeared which deals with the
topic of spurions in moose models in great detail.}, that are
external fields which for our purposes can be thought of being
constant. Note that our spurion fields differ from those
introduced in \cite{Arkani-Hamed:2001nc}. Their transformation
properties are assigned as
\begin{equation} q_L^p \rightarrow e^{i \xi_L^p}q_L^p e^{-i(\xi_L^p+\xi_R^p)/2}\quad \mathrm{and}\quad q_R^p \rightarrow e^{i
\xi_R^p}q_R^p e^{-i(\xi_L^p+\xi_R^p)/2} . \end{equation}
With their aid one can construct a Lagrangian that is chirally
invariant. The new covariant derivative is chosen such that it
transforms correctly under chiral transformations
\begin{equation} D_\mu U_p=\partial_\mu U_p -i q_L^p A^p_\mu q_L^{p\dagger}
U_p+ i U_p q_R^{p+1}A^{p+1}_\mu q_R^{p+1\dagger}
  \, .
\end{equation}
At the end of the calculation, the spurions will be set to
constant values $q_L,q_R\rightarrow \sqrt{g}\ \mathbbm{1}$,
therefore explicitly breaking the chiral symmetry. In order to
have a consistent assignment for the chiral dimension of the
covariant derivative, we choose the spurions to be of order
$\mathcal{O}(p^{1/2})$. This means that we count the gauge
coupling $g$ as $\mathcal{O}(p)$, similarly to \cite{Urech:1995hd}
which considered chiral perturbation theory with virtual photons.
Contributions of higher order in the coupling constant $g$ will
therefore be automatically suppressed at low energies.

Making use of the spurions, the requirement of $CP$-invariance and
translational invariance around the circle, the lowest order
Lagrangian $\mathcal{L}_2$, that is chirally invariant, contains
only the two terms written in Eq.(\ref{lag2_kin}), apart from
gauge-fixing and ghost terms.

Since we will study in this paper only one-loop corrections to the
masses of the gauge and Goldstone bosons, it will be sufficient to
consider only the following four terms in $\mathcal{L}_4$ (the
spurions have already been replaced by factors of $\sqrt{g}\
\mathbbm{1}$)
\begin{eqnarray} \label{eq:l4terms}
 \mathcal{L}_4 & = & c_1 g^2 v^2\!\sum_p \mathrm{Tr}\big[U_p^\dagger \big(D_\mu
 U_p\big) \, U_{p+1} \big(D^\mu U_{p+1}\big)^\dagger \big]+\frac{c_2
   g^2}{2}\!\sum_p
 \mathrm{Tr}\big[F_p^{\mu\nu} U_p F_{\mu\nu}^{p+1}U_p^\dagger\big] \nonumber \\
& & +c_3 g^2
 v^2\!\sum_p \mathrm{Tr}\big[D_\mu U_p \big(D^\mu U_p\big)^\dagger \big]+c_4
g^2\!\sum_p  \mathrm{Tr}\big[F_{\mu\nu}^p F_p^{\mu\nu}\big] \cdots
\ . \end{eqnarray}
For details see Appendix~\ref{sec:o4lag}. A priori, there is
another term of the form
\begin{equation} \left| \mathrm{Tr} \left[ q_R^1q_L^{1\dagger}U_1
      q_R^2q_L^{2\dagger}U_2\cdots U_N
  \right] \right|^2
\end{equation}
that is of $\mathcal{O}(p^{2N})$ and can potentially contribute.
Since we are particulary interested in the large $N$-limit, we
will choose $N \geq 4$, so that this term will not show up at
$\mathcal{O}(p^4)$.

\section{One-loop corrections to the gauge and Goldstone boson mass}
\label{sec:amumass}

 In 5D gauge theory compactified on a circle
the one-loop corrections to the Kaluza-Klein gauge-boson masses
have been calculated by several authors
\cite{vonGersdorff:2002as,Cheng:2002iz,Puchwein:2003jq,DaRold:2003yi}.
In the corresponding uncompactified theory, which is obtained in
the limit of the compactification radius $R$ going to infinity,
the 5D gauge bosons are massless at the classical level ($p^\mu
p_\mu\!=p_5^2$) due to gauge and Lorentz invariance. Since this
theory is non-renormalizable, radiative corrections can only be
calculated by introducing a cutoff. If one introduces this cutoff
in a way that respects gauge and Lorentz invariance, these
symmetries ensure that the gauge boson remains massless even at
the one-loop level. Since Lorentz invariance is violated in the
infrared by the compactification, finite mass corrections (that
vanish as $R\!\rightarrow\!\infty$) can be generated by radiative
corrections. Here we want to investigate this issue in the context
of deconstruction, which provides a manifestly gauge invariant
regularization for the above mentioned theory. The corrections to
the gauge boson masses at one-loop order are calculated for large
but finite $N$ and we show how to relate the results to the 5D
theory. \FIGURE[ht]{
\includegraphics[scale=1]{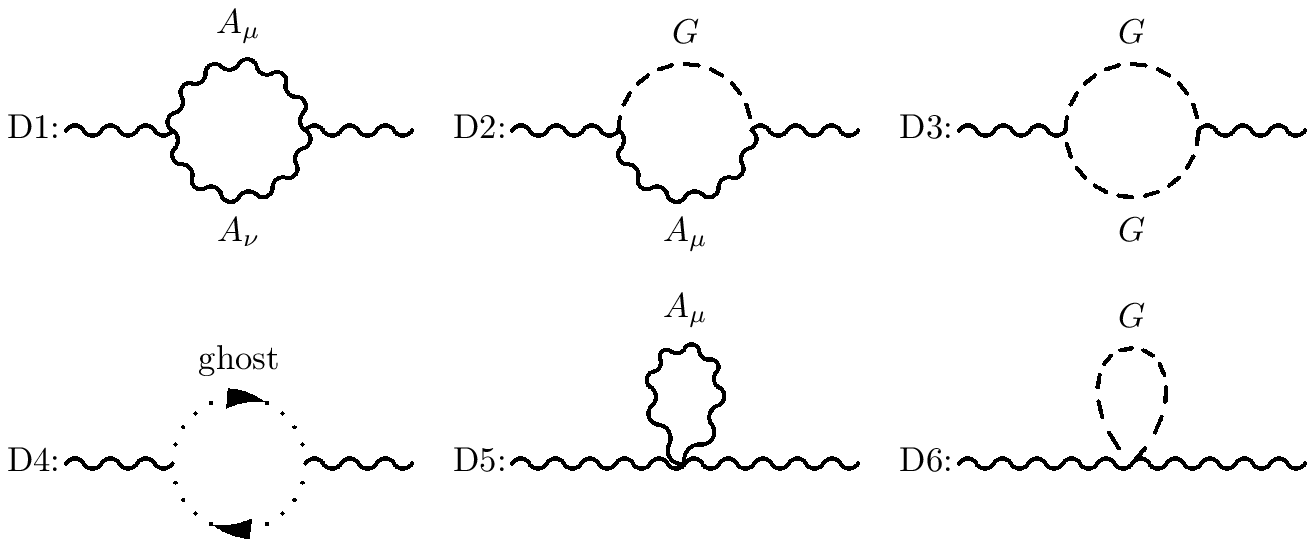}
\caption{Diagrams contributing to the $A_\mu$ mass.}
\label{fig:diags1} } We will find it advantageous to work in
Feynman gauge ($\xi=1$) where all fields have the same tree-level
mass spectrum. Using the Feynman rules from $\mathcal{L}_2$ and
the basis of wavefunctions from Appendix~\ref{sec:fourier} one can
work out the contributions to $\Pi_{\mu\nu}$ from all six one-loop
diagrams that contribute (see Figure~\ref{fig:diags1}). For the
calculation of the mass correction one is only interested in those
terms that are proportional to $g_{\mu\nu}$.
\begin{equation}
\Pi_{\mu\nu}(q)=g_{\mu\nu} \Pi_1(q^2) -q_\mu q_\nu \Pi_2(q^2)\ ,\qquad \lim_{q^2\rightarrow m_n^2} \Pi_1(q^2) \simeq
\delta m_n^2\ . \label{eq:deltm}
\end{equation}
The last relation in Eq.(\ref{eq:deltm}) is valid in the leading
order approximation. Details on the calculation and the results
for each individual diagram in Figure~\ref{fig:diags1} can be
found in Appendix~\ref{sec:diags}. Quite remarkably, using the
formulae given there, it is possible to work out the sums in a
closed form. However here we choose to expand in terms of $N^{-1}$
to simplify the result. The contribution from all one-loop
diagrams to $\delta m_n^2$ is (for $n>0$)
\begin{eqnarray}
\delta m_n^2\big|_{\mathrm{loops}} & = & \frac{g^2
  C_2(G)}{\pi^2}\Bigg\{\frac{m_n^2}{8}  \left(1+\frac{1}{12} \sin^2 \frac{n
\pi}{N}\right)\left[\frac{2}{\epsilon}
-\gamma +\log 4 \pi -\log \frac{g^2 v^2}{\mu^2}\right] \nonumber \\
& & +\left[ \frac{3}{4} \frac{g^2 v^2}{ N^3} \zeta(3) - \frac{15}{4}
\frac{g^2 v^2}{N^5}\zeta
(5)+\mathcal{O}\left(\frac{1}{N^7}\right)\right] -
\left[\frac{7}{96}-\frac{3}{8 N^3}
\zeta(3)+\mathcal{O}\left(\frac{1}{N^5}\right)\right]m_n^2 \nonumber \\
& &+
\left[\frac{11}{96}+\mathcal{O}\left(\frac{1}{N^3}\right)\right]m_n^2
\sin^2 \frac{n \pi}{N} + m_n^2\mathcal{O}\left( \sin^4 \frac{n
\pi}{N}\right)\Bigg\}\ . \label{eq:deltmn}
 \end{eqnarray}
Apart from the corrections coming from loop diagrams, we also have
to include contributions from the terms in $\mathcal{L}_4$. Due to
their higher order in $p$ they enter only at tree level. Their
contribution to the gauge-boson mass originates from the four
terms in Eq.(\ref{eq:l4terms})
\begin{equation}
\delta m_n^2\big|_{\mathcal{L}_4}=g^2 \mu^\epsilon m_n^2
\Big((c_1+c_2+c_3+c_4)-2(c_1+c_2)\sin^2\frac{n \pi}{N}\Big)\
.\label{eq:counter}
\end{equation}
The $1/\epsilon$ singularities appearing in Eq.(\ref{eq:deltmn})
are cancelled by the divergent part of the $c_i$ (see
Eq.(\ref{eq:loenco})). The renormalized parameters
$c_i^\mathrm{r}$ are coupling constants of the non-linear
$\sigma$-model. Since mass corrections proportional to $m_n^2$ and
$m_n^2 \sin^2 n\pi/N$ involve the constants $c_i^\mathrm{r}$ they
can only be predicted if the $c_i^\mathrm{r}$ can be extracted by
fits to additional experimental data\footnote{Because the
constants $c_i^\mathrm{r}$ enter only in the linear combinations
$c_1^\mathrm{r}+c_2^\mathrm{r}$ and
$c_3^\mathrm{r}+c_4^\mathrm{r}$, to determine all of them, it is
not sufficient to measure the mass splittings only, but one also
has to consider other processes such as scattering amplitudes.}.
If one makes a linear completion of this model these constants can
be calculated in the context of a linear sigma model
\cite{Falkowski:2003iy}. They will however depend on the details
of the linearization such as the choice of the Higgs-potential.

The question remains how the 5D limit is obtained at one-loop
order. To avoid the subtleties of the compactification, we will
consider the limit of $N\!\rightarrow\!\infty$ which corresponds
to $R\!\rightarrow\!\infty$ in the 5D theory. In this limit the 5D
theory possesses full 5D Lorentz invariance. However since we are
working in a latticized model we expect this symmetry to be
violated at scales smaller than the lattice spacing $a$. When
going to the uncompactified theory the fifth component of the
momentum becomes a continuous variable (as before we have $2\pi
R=N a$)
\begin{equation}
p_5^2=\frac{n^2}{R^2}\ .
\end{equation}
The corresponding relation for the mass $m_n^2$ reads
\begin{equation}
m_n^2=4 g^2 v^2 \sin^2 \frac{n \pi}{N}= 4 g^2 v^2 \sin^2 \frac{p_5
a}{2}\ .
\end{equation}
We require that the violation of the Lorentz symmetry does not
occur at distances larger than the lattice spacing $a$. Therefore
we demand that
\begin{equation}
\label{eq:lorsym}
 p_5^2(1 +\mathcal{O}(p_5
a))=\lim_{N\rightarrow\infty}(m_n^2+\delta m_n^2)\ .
\end{equation}
Combining the one-loop corrections
Eqs.(\ref{eq:deltmn},\ref{eq:counter}) with the tree level
expression we find that the mass in the limit $N\rightarrow\infty$
is given by
\begin{equation}
\lim_{N\rightarrow\infty}(m_n^2+\delta m_n^2)=4 g^2 v^2
\sin^2\frac{p_5 a}{2}\left[1\!+\!g^2\bigg(\sum_{i=1}^4
c_i^{\mathrm{r}}(\mu)-\!\frac{C_2(G)}{8
\pi^2}\bigg(\!\frac{7}{12}\!+\!\log \frac{g^2
v^2}{\mu^2}\bigg)\bigg)\right]\ .
\end{equation}
In order for Eq.(\ref{eq:lorsym}) to be fulfilled, the tree level
relation for the lattice spacing in terms of parameters of the 4D
EFT (Eq.(\ref{eq:treeid})) has to be modified by quantum
corrections, as has been pointed out in
\cite{Arkani-Hamed:2001ca}. The properly chosen lattice spacing at
one-loop order, that fulfills Eq.(\ref{eq:lorsym}) is given by
\begin{equation}
a^2_{(1)}=\frac{1}{v^2 g^2}\left[1\!-\!g^2\bigg(\sum_{i=1}^4
c_i^{\mathrm{r}}(\mu)-\!\frac{C_2(G)}{8
\pi^2}\bigg(\!\frac{7}{12}\!+\!\log \frac{g^2
v^2}{\mu^2}\bigg)\bigg)\right]\ .
\end{equation}
Note that if $m_n^2+\delta m_n^2$ is $\mu$-independent, $a_{(1)}$
is also independent of $\mu$. With the appropriate definition of
$a$ the mass corrections are well defined in the limit
$a\rightarrow 0$. Keeping the radius $R$ fixed, one recovers the
limit of a continuous and compactified fifth dimension.
Introducing the 5D gauge coupling $g_5^2=a g^2$ one finds that the
mass corrections are given by
\begin{equation}
\label{eq:ares} m_n^2+\delta m_n^2 \rightarrow \frac{n^2}{R^2}
+\frac{3}{4} \frac{g^2_5}{2 \pi R} \frac{C_2(G) \zeta(3)}{4 \pi^4
R^2}\ .
\end{equation}
This is exactly the result obtained in the 5D calculations
\cite{Cheng:2002iz,Puchwein:2003jq}. Having worked in a manifestly
gauge invariant framework and having maintained 5D Lorentz
invariance in the $R\rightarrow\infty$ limit by the appropriate
choice of the lattice spacing, we have obtained a  finite
result\footnote{This has already been pointed out, but not
explicitly been shown in \cite{Puchwein:2003jq}.} as was discussed
in the beginning of Section~\ref{sec:amumass}.

We have not yet dealt with the zero mode. Since all contributions
from $\mathcal{L}_4$ are proportional to $m_n^2$, which is zero in
this case, they do not appear in the mass correction of the zero
mode. The explicit calculation shows that the correction is in
fact exactly zero, as all diagrams in Figure~\ref{fig:diags1}
cancel each other. This is consistent with the 5D calculation,
since in the limit of a continuous dimension the zero-mode mass
has to vanish because of 5D gauge symmetry.

\FIGURE[ht]{
\includegraphics[scale=1]{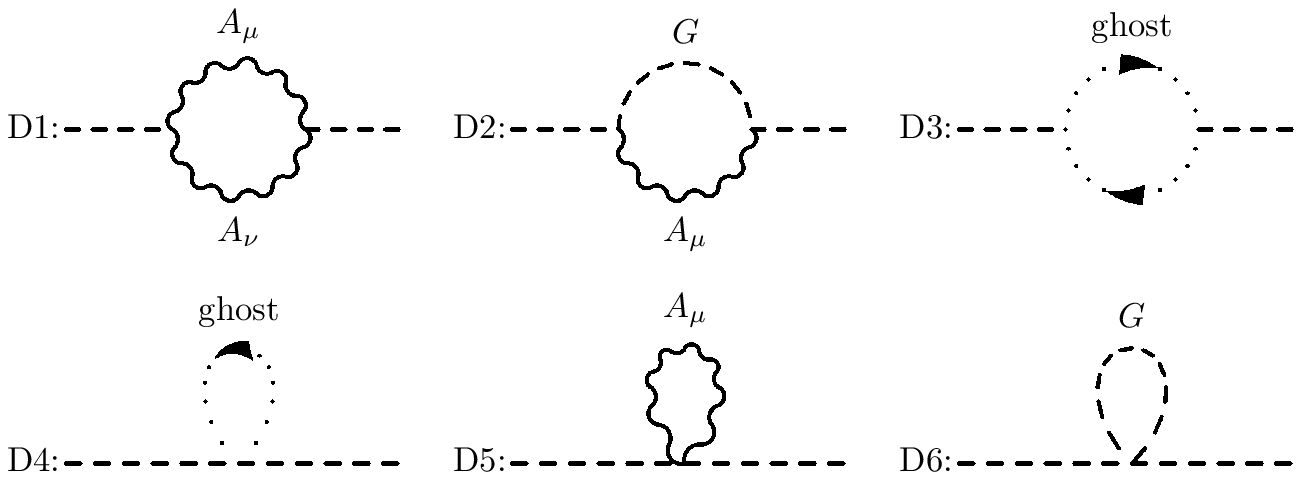}
\caption{Diagrams contributing to the Goldstone-boson mass}
\label{fig:diags2} } The situation for the mass of the
Goldstone-boson is somewhat different from the case of the
$A_\mu$. Going to unitary gauge one can show that the only
physical mode is the zero mode $G_0$. All other modes are eaten
due to the breakdown of the gauge symmetry to the diagonal
subgroup ($SU(k)^N\! \rightarrow\! SU(k)_{\mathrm{diag.}}$) and
form the heavy gauge fields \cite{Arkani-Hamed:2001nc}. Similarly
to the case of the $A_\mu$ zero-mode the mass corrections have to
be finite due to the absence of terms in $\mathcal{L}_4$ that can
absorb possible $1/\epsilon$-poles. However in this case the
correction is not zero. The sum of diagrams from figure
Figure~\ref{fig:diags2} becomes
\begin{equation}
\label{eq:massgold} \delta m_0^2 = \frac{g^2 C_2(G)}{\pi^2}\left[
\frac{9}{4} \frac{g^2 v^2}{ N^3} \zeta(3) + \frac{45}{4} \frac{g^2
v^2}{N^5}\zeta (5)+\mathcal{O}\left(\frac{1}{N^7}\right)\right]\ .
\end{equation}
In the limit of going over to a continuous fifth dimension the
shift is three times what we have found for the $A_\mu$-mass
(Eq.(\ref{eq:ares})).

\section{Conclusion}
When doing calculations in 5D theories one is immediately
confronted with the problem of non-renormalizability.
Deconstruction is one way of introducing a gauge invariant
UV-regulator in the theory, similar to conventional lattice
regularization. Although the non-linear $\sigma$-model obtained by
this procedure as a low energy EFT is non-renormalizable, one can
apply the methods from chiral perturbation theory to obtain
sensible and finite results at every order in the low energy
expansion. We have illustrated how to apply these methods to the
case of deconstruction and have given an explicit example by
calculating the corrections to the masses of the gauge  boson and
the Goldstone boson in a non-abelian pure-gauge theory. In order
to retain the correspondence between the 5D and the deconstructed
theory at higher order, a sensible definition of the lattice
spacing is necessary at a given order in perturbation theory to
ensure 5D Lorentz invariance. With the appropriate definition the
gauge and Goldstone boson masses become independent of the cutoff
and the continuum limit can be taken. In this limit the one-loop
mass corrections are finite due to 5D gauge and Lorentz invariance
and agree with earlier 5D calculations
\cite{vonGersdorff:2002as,Cheng:2002iz,Puchwein:2003jq,DaRold:2003yi}.
We have demonstrated that deconstruction, in the cases
investigated in this paper, can serve as a viable UV-completion
for 5D theories, even at the one-loop order.

\acknowledgments We thank A. Falkowski and S. Pokorski for
providing us with details on their related work on the linear
$\sigma$-model \cite{Falkowski:2003iy} and for discussions. This
research was supported  in part by the National Science Foundation
under the Grant No. PHY99-07949 and the Swiss National Science
Foundation under Grant No. SNF20-68037.02. Z. K. and M. P. also
thank the Kavli Institute for Theoretical Physics, UCSB for
hospitality during the completion of this work.

\appendix
\section{Mode decomposition}
\label{sec:fourier} We first introduce a basis of functions on the set of points $p=1,\ldots,N$ that fulfills the
requirement of $N$-periodicity. The functions
\begin{align}
&g^+_n(p)=\sqrt{\frac{2}{N}} \cos \frac{2 n \pi p}{N}\ ,\quad&g^-_n(p)&=\sqrt{\frac{2}{N}} \sin \frac{2 n \pi p}{N}\ ,\nonumber\\
&g^+_0(p)=\frac{1}{\sqrt{N}}\ ,&g^+_{N/2}(p)&=\frac{1}{\sqrt{N}}(-1)^p\ ,&& \mathrm{with}\ n\in\{1,\ldots ,N/2-1\}\ ,
\end{align}
form a complete set and serve as an orthonormal basis with respect
to the scalar product ($\sigma,\sigma'\in\{+,-\}$)
\begin{equation}
\label{eq:gnorm}
 \sum_{p=1}^N g_n^\sigma(p)
g_{n'}^{\sigma'}(p)=\delta_{n n'}\delta_{\sigma \sigma'}\ .
\end{equation}
Expanded in this basis the gauge- and the ghost-field read
\begin{equation}
A_\mu^p=\sum_n\Big(A_\mu^{+,n} g^+_n(p)+A_\mu^{-,n} g^-_n(p)\Big)\ ,\qquad c_p=\sum_n\Big(c^+_n g^+_n(p)+c^-_n
g^-_n(p)\Big)\ .\label{eq:ac_decomp}
\end{equation}
Since the Goldstone boson fields $G_p$ are associated with the
link between the points $p$ and $p+1$ we choose a different
Fourier basis
\begin{align}
&f^+_n(p)=\sqrt{\frac{2}{N}}
\cos \frac{2 n \pi (p+\frac{1}{2})}{N}\ ,&f^-_n(p)&=-\sqrt{\frac{2}{N}} \sin \frac{2 n \pi (p+\frac{1}{2})}{N}\ ,\nonumber\\
&f^+_0(p)=\frac{1}{\sqrt{N}}\
,&f^-_{N/2}(p)&=\frac{1}{\sqrt{N}}(-1)^{p+1}\ ,\quad
\mathrm{with}\ n\in\{1,\ldots, N/2-1\}\ .
\end{align}
The corresponding expansion for the Goldstone bosons reads
\begin{equation}
G_p=\sum_n \Big(G^+_n f^+_n(p)+G^-_n f^-_n(p)\Big)\
.\label{eq:g_decomp}
\end{equation}
In this basis the free part of the Lagrangian $\mathcal{L}_2$
becomes diagonal and we obtain the propagators
\begin{align}
\label{eq:props}
 &(D_{\mathrm{GB}}^{ab})^{\sigma\sigma'}_{n
n'}(k)=\frac{i}{k^2-\xi m_n^2+i \epsilon}
\delta^{\sigma\sigma'}\delta_{n n'}\delta^{ab}\ ,\quad
(D_{\mathrm{ghost}}^{ab})^{\sigma\sigma'}_{n
n'}(k)=\frac{i}{k^2-\xi m_n^2+i \epsilon}
\delta^{\sigma\sigma'}\delta_{n n'}\delta^{ab}\ ,\nonumber \\
&(D_{\mathrm{gauge}}^{\mu\nu,ab})^{\sigma\sigma'}_{n
n'}(k)=-i\Big[\frac{g^{\mu\nu}}{k^2-m_n^2+i
\epsilon}-(1-\xi)\frac{k^\mu k^\nu}{(k^2-\xi m_n^2+i \epsilon)
(k^2-m_n^2+i \epsilon)}\Big]\delta^{\sigma\sigma'}\delta_{n
n'}\delta^{ab}\ .
\end{align}

\section{Lagrangian at $\mathcal{O}(p^4)$}
\label{sec:o4lag} For the purpose of this paper, where we study
only the one-loop corrections to the masses of the gauge bosons
and the Goldstone boson, it is not necessary to construct a
complete list of operators at $\mathcal{O}(p^4)$ and then reduce
it to a minimal set of independent operators, using algebraic
identities and the equations of motion. In particular, the issue
of preserving gauge invariance when using the equations of motion
for the Goldstone and the gauge bosons is not so straightforward
as in ordinary chiral perturbation theory in the case of local
gauge symmetries, see \cite{Nyffeler:1999ap}. Furthermore, since
we do not consider other observables, like scattering amplitudes,
we will not be able to extract for instance the divergent terms of
individual low-energy constants $c_i$ (Eq.(\ref{eq:loenco})).
Therefore, we give below only some counter terms that are needed
to render the masses of the gauge bosons finite. In particular,
one finds the following four terms that give a contribution to the
mass of $A_\mu$, see also \cite{Hirn:2004ze},
\begin{align}
c_1 v^2 &\sum_p \mathrm{Tr}\big[U_p^\dagger D_\mu U_p \, q_R^{p+1}
q_L^{p+1\dagger} \, U_{p+1} \big(D^\mu U_{p+1}\big)^\dagger \,
q_L^{p+1} q_R^{p+1\dagger}
\big]\ ,\nonumber\\
\frac{c_2}{2} &\sum_p \mathrm{Tr}\big[q_L^p F_p^{\mu\nu}
q_L^{p\dagger} \, U_p \, q_R^{p+1} F_{\mu\nu}^{p+1}
q_R^{p+1\dagger}
U_p^\dagger\big]\ ,\nonumber\\
c_3 g^2 v^2 &\sum_p \mathrm{Tr}\big[D_\mu U_p
(D^\mu U_p)^\dagger \big]\ ,\nonumber\\
c_4 g^2 &\sum_p \mathrm{Tr}\big[F_{\mu\nu}^p F^{\mu\nu}_p \big]\
.\label{eq:l4terms2}
\end{align}
We have written the last two terms in a shorthand notation that
comprises a set of terms that differ by various insertion of
spurions. After setting the spurions to $\sqrt{g}\ \mathbbm{1}$
they give the same contribution, which is proportional to the
corresponding terms in the lowest order Lagrangian
$\mathcal{L}_2$. One can convince oneself that all terms in
Eq.(\ref{eq:l4terms2}) give contributions to the mass correction
that are proportional to $m_n^2$ or $m_n^2
\sin^2(n\pi/N)$.~\footnote{To obtain
  higher powers in $\sin(n\pi/N)$ one would need to construct terms that
  involve products of fields at distant points $p$ (e.g. $\mathrm{Tr}\big[q_L^{p}
  F_{\mu\nu}^p q_L^{p\dagger} \, U_p \, q_R^{p+1} q_L^{p+1\dagger} \, U_{p+1}
  \, q_R^{p+2} F^{\mu\nu}_{p+2} q_R^{p+2\dagger} \, U_{p+1}^\dagger \,
  q_L^{p+1} q_R^{p+1\dagger} \, U_p^\dagger \big]$ for contributions of the
  form $m_n^2 \sin^4(n\pi/N)$),  but these terms are of higher power in
  momentum due to the spurion fields.}  The same is true for the Goldstone
boson where $m_n^2 = 0$.  Therefore no counter term is needed at
$\mathcal{O}(p^4)$ for the corrections to the mass of the
Goldstone boson and the zero-mode of the gauge boson.

\section{Diagrams contributing at one-loop order}
\label{sec:diags}
 \TABLE[ht]{
\includegraphics[scale=1]{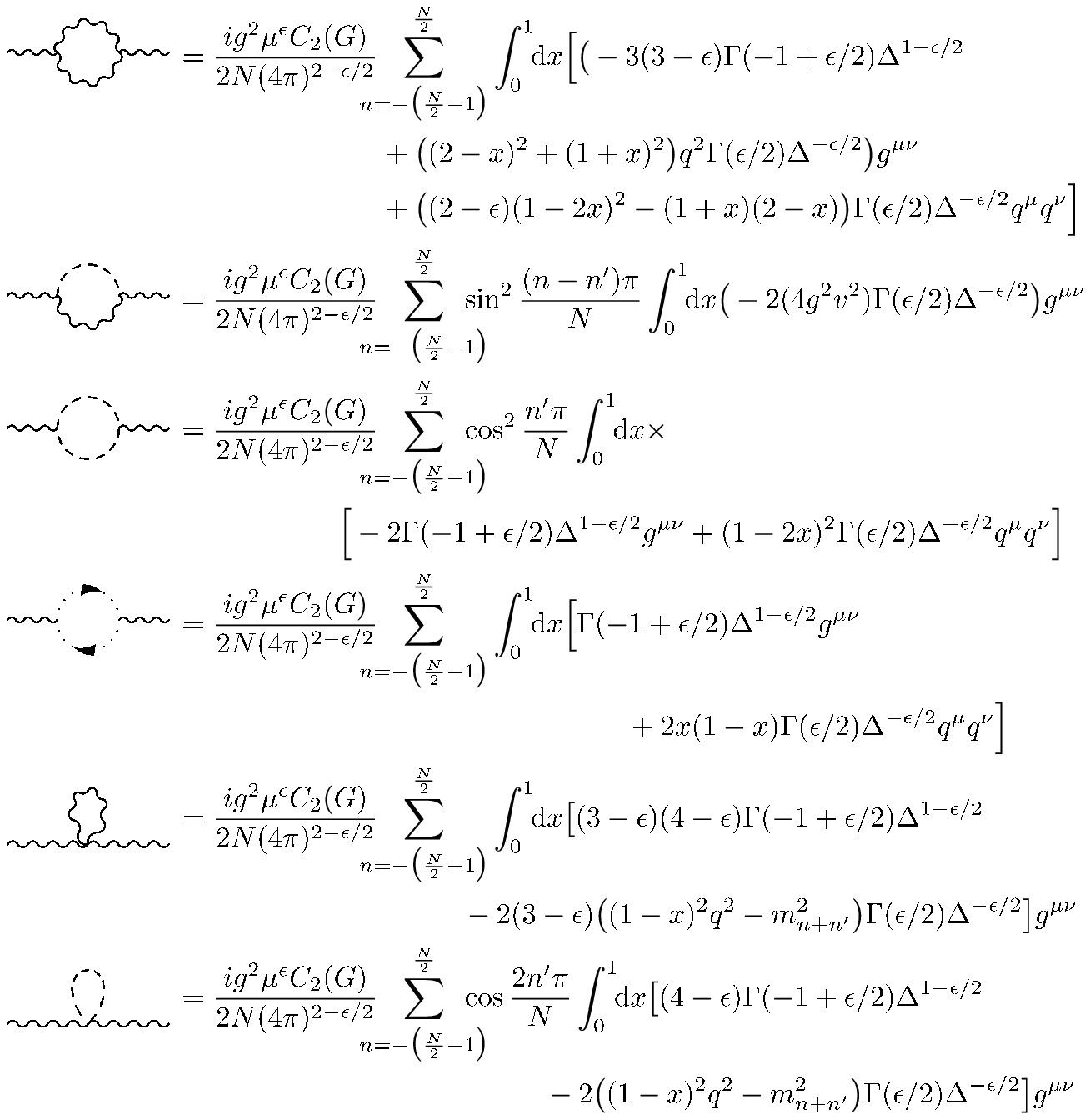}
\caption{Corrections to the gauge boson propagator
($\Delta=(1-x)m_n^2+ x m_{n+n'} - x(1-x) q^2$ and $n'$ labels the
external states).} \label{fig:a_mu_res} }
 In this section we give details on the calculation of all one-loop diagrams
that contribute to the gauge boson self-energy
(Figure~\ref{fig:diags1}) in next-to-leading order. The first step
is to obtain the Feynman rules. One expands the fields in terms of
their Fourier modes Eqs.(\ref{eq:ac_decomp},\ref{eq:g_decomp}).
Making use of trigonometric identities and using the
orthonormality relations Eq.(\ref{eq:gnorm}) one can evaluate the
sum over $p$ and read off the vertex functions for the individual
vertices. As an example we consider the case of the three-vertex
of two gauge and one Goldstone boson field. For the special choice
of parities given in Figure~\ref{fig:vert} the vertex function
becomes
\begin{eqnarray}
V_{A A G}^{+ + +}&=&i\frac{g^2 v^2 N}{4}
f^{abc}\Big[\sin\frac{(2n_1-n_3)\pi}{N}
\big(\tilde{\delta}_{n1-n2,n3}+\tilde{\delta}_{n1+n2,n3}\big)\nonumber\\
&=&-
\sin\frac{(2n_1+n_3)\pi}{N}\big(\tilde{\delta}_{n1-n2,-n3}+\tilde{\delta}_{n1+n2,-n3}\big)\Big]
\eta_{n_1} \eta_{n_2} \eta_{n_3} g^{\mu\nu}
\end{eqnarray}
where
\begin{equation}
\tilde{\delta}_{n,n'}=\bigg\{\begin{array}{l}
  1\ \mathrm{if}\ (n-n')\!\!\!\!\!\mod N=0 \\
  0\ \mathrm{else} \\
\end{array}\ ,\qquad
\eta_n=\bigg\{\begin{array}{l}
  \sqrt{1/N}\ \mathrm{if}\ n=0,N/2 \\
  \sqrt{2/N}\ \mathrm{else} \\
\end{array}\ .
\end{equation}
\FIGURE[htb]{\includegraphics[scale=1]{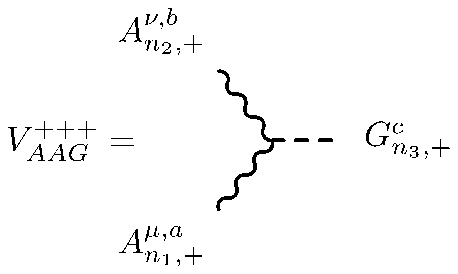} \caption{Two
gauge, one Goldstone boson vertex.}\label{fig:vert}} The remaining
vertex functions can be obtained in a similar way. With this
information at hand, the expressions for all diagrams that
contribute to the one-loop gauge boson self energy can be written
down using the propagators from Eq.(\ref{eq:props}). One is left
with two sums over Kaluza-Klein labels of the propagating internal
states, one of which can be evaluated by making use of the
Kronecker delta. Because of translation invariance the diagrams
are only non-vanishing when both external states have the same
Kaluza-Klein label. One proceeds in the usual way by Feynman
parameterizing  the integrals and evaluating them in dimensional
regularization. The results are given in Table \ref{fig:a_mu_res}.
For the calculation of the mass corrections we are interested in
the on-shell limit of these expressions. Quite remarkably the
remaining sums can be evaluated in a closed from with the aid of
the following two relations (here $\psi_0$ denotes the
Digamma-function)\footnote{Eq.(\ref{eq:sum2}) can be derived by
the use of Gauss's Digamma theorem \cite{Erdelyi:1953}.}.
\TABLE[ht]{
\includegraphics[scale=1]{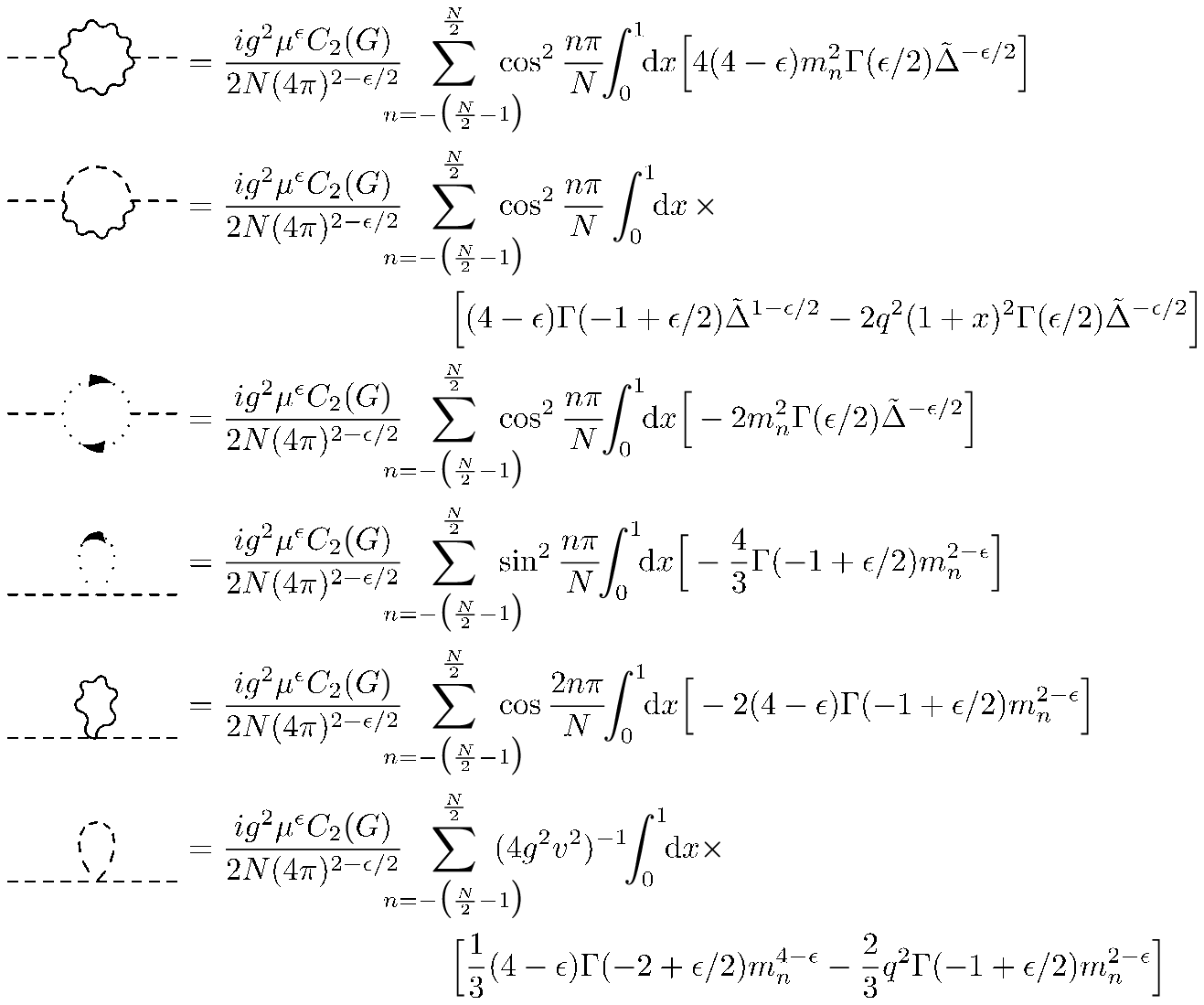}
\caption{Corrections to the Goldstone boson propagator
($\tilde\Delta=m_n^2 - x(1-x) q^2$).} \label{fig:g_res} }
\begin{align}
&\sum_{n=-(N/2-1)}^{N/2}\!\!\!\!\sin^{2m} \left( \frac{n \pi}{N}
\right) =\frac{N}{2^{2m}} \bigg(\begin{matrix}
  2m \\
  m \\
\end{matrix}\bigg)\ ,\\
&\sum_{n=-(N/2-1)}^{N/2}\!\!\!\!\sin^{2m} \left( \frac{n \pi}{N}
\right) \log
\left[ \sin^2 \frac{n \pi}{N}\right]\nonumber\\
&\qquad=\frac{1}{2^{2m-1}}\left[ \sum_{k=1}^m (-1)^k
\bigg(\begin{matrix}
  2m \\
  m\!-\!k \\
\end{matrix}\bigg)
\big(\psi_0(k/N)\!+\!\psi_0(1\!+\!k/N)\big)-\!\bigg(\begin{matrix}
  2m \\
  m \\
\end{matrix}\bigg)
\left(\gamma\!+\!N \log 2\right)\right]\,.\label{eq:sum2}
\end{align}
The mass correction Eq.(\ref{eq:deltmn}) follows from expanding
the rather lengthly result in the case of large $N$. In the
special case of the zero-mode with $m_{n'}^2=0$ the correction is
found to be vanishing.

Next we want to do the corresponding calculation for the Goldstone
boson. Here the situation is somewhat easier since we only have to
deal with the zero-mode, being the only physical degree of
freedom. Proceeding as before the contributing diagrams from
Figure~\ref{fig:diags2} give the corrections listed in Table
\ref{fig:g_res}. After evaluating the sums and expanding in
$N^{-1}$ the mass correction is the one given in
Eq.(\ref{eq:massgold}).

\bibliographystyle{JHEP}
\bibliography{paper}
\end{document}